\documentclass[nonacm,acmsmall,screen]{acmart}

\makeatletter
\let\wfs@comment@comment\comment
\let\comment\@undefined
\makeatother

\usepackage{marvosym}
\usepackage{algorithmic}
\usepackage{graphicx}
\usepackage{textcomp}
\usepackage{xcolor}
\usepackage{float}
\usepackage[capitalize,nameinlink]{cleveref}
\usepackage{pdfcomment}
\usepackage{tikzit}

\Crefformat{equation}{#2Graph~#1#3}

\hypersetup{
    keeppdfinfo
}

\definecolor{orange}{RGB}{200,127,0}
\definecolor{yellowish}{RGB}{235,207,52}

\newcommand{\comment}[2][]{\pdfmargincomment[icon=Comment,author=#1]{#2}}

\begin{document}

\title{HUGR: A Quantum-Classical Intermediate Representation}

\author{Mark Koch}
\orcid{0000-0001-8511-2703}
\author{Agustín Borgna}
\orcid{0000-0002-1688-1370}
\author{Seyon Sivarajah}
\orcid{0000-0002-7332-5485}

\author{Alan Lawrence}
\orcid{0009-0000-1663-7397}
\author{Alec Edgington}
\orcid{0000-0002-0508-6988}
\author{Douglas Wilson}

\author{Craig Roy}
\orcid{0009-0002-6034-2910}
\author{Luca Mondada}
\orcid{0000-0002-7496-7711}
\author{Lukas Heidemann}
\orcid{0000-0002-7137-2368}

\author{Ross Duncan}
\orcid{0000-0001-6758-1573}

\affiliation{%
\institution{Quantinuum}
\city{Cambridge}
\country{United Kingdom}
}


\begin{abstract}
We introduce the Hierarchical Unified Graph Representation (HUGR): a novel graph based intermediate representation for mixed quantum-classical programs. HUGR’s design features high expressivity and extensibility to capture the capabilities of near-term and forthcoming quantum computing devices, as well as new and evolving abstractions from novel quantum programming paradigms. The graph based structure is machine-friendly and supports powerful pattern matching based compilation techniques. Inspired by MLIR, HUGR’s extensibility further allows compilation tooling to reason about programs at multiple levels of abstraction, lowering smoothly between them. Safety guarantees in the structure including strict, static typing and linear quantum types allow rapid development of compilation tooling without fear of program invalidation. A full specification of HUGR and reference implementation are open-source and available online.
\end{abstract}

\hypersetup{urlcolor=black}
    \maketitle
\hypersetup{urlcolor=ACMDarkBlue}

\pagestyle{plain}


\section{Introduction}%
\label{sec:introduction}

Modern applications of quantum computers usually involve both quantum and
classical processors interacting with each other. In particular, there is an
increasing interest in algorithms that require classical decision making
\emph{during} the execution of a circuit (i.e. within the coherence time of its
qubits). For example, recently demonstrated repeat-until-success
protocols~\cite{rus-paetznick2013,brown2023advancescompilationquantumhardware} and other
algorithms use classical control-flow conditioned on
mid-circuit measurements to determine which quantum operations should be applied
next~\cite{Griffiths1996Semiclassical, Kuperberg2005Subgroup, Ozols2013RejectionSampling}.
Going beyond that, quantum error correction algorithms, for example, might
require even more complex classical logic to decode errors in real-time and
apply corrections to the quantum state~\cite{Shor1995QEC,Stean1996QEC}.

Enabling this tight integration between the quantum and classical processor
requires dedicated support throughout the entire quantum software stack. In
particular, as previously argued 
in~\cite{brown2023advancescompilationquantumhardware,qiro,McCaskey2021MlirDialect}, 
there is a need for an
intermediate representation (IR) for quantum programs that natively captures
these classical operations, going beyond the traditional circuit picture. To
this end, we introduce the Hierarchical Unified Graph Representation (HUGR), a
novel quantum IR that can efficiently express, reason about, and optimise these
hybrid quantum-classical programs in a unified graph structure. Its design is
guided by the following main principles:

\paragraph{Expressivity}
HUGR captures the various computational requirements of quantum algorithms in
one unified framework. It can express everything from traditional, static
(possibly parameterised) circuits and hybrid quantum-classical optimisation
loops up to the real-time quantum classical logic described above.

\paragraph{Machine-friendliness}
As an intermediate representation, HUGR is designed to be efficiently consumable
and manipulable by software. We do not expect end users to read or write HUGR
directly. Instead, they should rely on front-ends like higher-level programming
languages or libraries that compile to HUGR.

\paragraph{Abstraction}
Quantum algorithms are usually built up from multiple layers of abstraction, starting from some problem domain (say, a Hamiltonian describing a chemical system) that is then continuously lowered (for example by synthesising oracle circuits, inserting error mitigation steps, etc.).
HUGR is designed to faithfully capture this staged lowering of abstraction levels, allowing compilers to exploit the unique opportunities for optimisation available at each step.

\paragraph{Extensibility}
HUGR follows a modular design where new operations and data types can be added
on the fly, comparable to the dialect system in the MLIR compiler tool
chain~\cite{mlir}. This allows third parties to define their own bespoke
abstractions and lowering routines that seamlessly compose with other components
and passes.

\paragraph{Optimisability}
HUGR is designed to enable efficient optimisation of quantum programs, both
within and across the quantum-classical boundary. On top of that, we provide
efficient routines for matching and rewriting of patterns within HUGR programs 
that third parties can hook into to define their own domain-specific optimisation 
routines (see \Cref{sec:optimisation} for details).

\medskip
The full specification and reference implementation of HUGR are open-source and available at \href{http://www.github.com/CQCL/hugr}{\texttt{github.com/CQCL/hugr}}.


\section{The Hierarchical Unified Graph Representation}%
\label{sec:hugr}

\subsection{Quantum Programs as Dataflow Graphs}%
\label{sec:hugr:structure}

Quantum compiler architectures usually represent quantum circuits in the form of
directed acyclic graphs (DAGs) where nodes are quantum gates and edges describe
the qubit dependencies between them. HUGR generalises this model by encoding
both quantum and classical operations in the same DAG structure. Concretely,
HUGR represents programs via dataflow graphs spanned between an input and output
node, where edges can carry either qubits or arbitrary classical data. The nodes
correspond to quantum or classical processes that act on these values and
produce some outputs that can be fed to the following nodes in the graph:
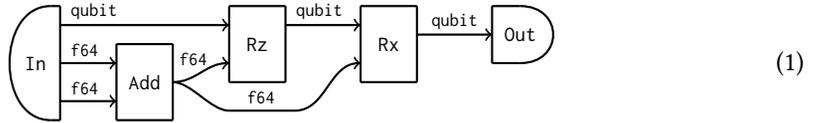
\begin{equation}%
    \label{eq:hugr:rz}
    \begin{tikzpicture}
	\begin{pgfonlayer}{nodelayer}
		\node [style=none] (2) at (-0.625, 0.75) {};
		\node [style=none] (3) at (-0.625, -0.75) {};
		\node [style=none] (4) at (-0.825, 0) {\footnotesize\texttt{In}};
		\node [style=none] (5) at (0.25, -0.75) {};
		\node [style=none] (6) at (1, -0.75) {};
		\node [style=none] (7) at (1, 0.25) {};
		\node [style=none] (8) at (0.25, 0.25) {};
		\node [style=none] (9) at (0.625, -0.25) {\footnotesize \texttt{Add}};
		\node [style=none] (10) at (-0.5, -0.5) {};
		\node [style=none] (11) at (0.25, -0.5) {};
		\node [style=none] (12) at (-0.5, 0) {};
		\node [style=none] (13) at (0.25, 0) {};
		\node [style=none] (14) at (-0.175, -0.325) {\scriptsize\texttt{f64}};
		\node [style=none] (15) at (-0.175, 0.175) {\scriptsize\texttt{f64}};
		\node [style=none] (16) at (-0.5, 0.5) {};
		\node [style=none] (18) at (-0.05, 0.675) {\scriptsize\texttt{qubit}};
		\node [style=none] (19) at (1.75, -0.25) {};
		\node [style=none] (20) at (2.5, -0.25) {};
		\node [style=none] (21) at (2.5, 0.75) {};
		\node [style=none] (22) at (1.75, 0.75) {};
		\node [style=none] (23) at (2.125, 0.25) {\footnotesize \texttt{Rz}};
		\node [style=none] (24) at (1.75, 0) {};
		\node [style=none] (25) at (1.75, 0.5) {};
		\node [style=none] (26) at (1, -0.25) {};
		\node [style=none] (27) at (2.5, 0.5) {};
		\node [style=none] (28) at (5.625, 0.75) {};
		\node [style=none] (29) at (5.625, 0) {};
		\node [style=none] (30) at (5.625, 0.375) {\footnotesize \texttt{Out}};
		\node [style=none] (31) at (5.25, 0.75) {};
		\node [style=none] (32) at (5.25, 0) {};
		\node [style=none] (34) at (3.5, -0.25) {};
		\node [style=none] (35) at (4.25, -0.25) {};
		\node [style=none] (36) at (4.25, 0.75) {};
		\node [style=none] (37) at (3.5, 0.75) {};
		\node [style=none] (38) at (3.875, 0.25) {\footnotesize \texttt{Rx}};
		\node [style=none] (39) at (3.5, 0) {};
		\node [style=none] (40) at (3.5, 0.5) {};
		\node [style=none] (41) at (4.25, 0.375) {};
		\node [style=none] (42) at (5.25, 0.375) {};
		\node [style=none] (43) at (1.75, -0.625) {};
		\node [style=none] (44) at (2.625, -0.625) {};
		\node [style=none] (45) at (-0.5, -0.75) {};
		\node [style=none] (46) at (-0.5, 0.75) {};
		\node [style=none] (47) at (2.95, 0.675) {\scriptsize\texttt{qubit}};
		\node [style=none] (48) at (2.175, -0.45) {\scriptsize\texttt{f64}};
		\node [style=none] (49) at (1.275, 0.05) {\scriptsize\texttt{f64}};
		\node [style=none] (50) at (4.75, 0.55) {\scriptsize\texttt{qubit}};
	\end{pgfonlayer}
	\begin{pgfonlayer}{edgelayer}
		\draw [thick, bend right=90, looseness=1.25] (2.center) to (3.center);
		\draw [thick] (8.center) to (5.center);
		\draw [thick] (5.center) to (6.center);
		\draw [thick] (6.center) to (7.center);
		\draw [thick] (7.center) to (8.center);
		\draw [thick, ->] (10.center) to (11.center);
		\draw [thick, ->] (12.center) to (13.center);
		\draw [thick] (22.center) to (19.center);
		\draw [thick] (19.center) to (20.center);
		\draw [thick] (20.center) to (21.center);
		\draw [thick] (21.center) to (22.center);
		\draw [thick, ->] (16.center) to (25.center);
		\draw [thick, ->, in=180, out=0] (26.center) to (24.center);
		\draw [thick, bend left=90, looseness=2.00] (28.center) to (29.center);
		\draw [thick] (32.center) to (29.center);
		\draw [thick] (32.center) to (31.center);
		\draw [thick] (31.center) to (28.center);
		\draw [thick] (37.center) to (34.center);
		\draw [thick] (34.center) to (35.center);
		\draw [thick] (35.center) to (36.center);
		\draw [thick] (36.center) to (37.center);
		\draw [thick, ->] (41.center) to (42.center);
		\draw [thick, ->] (27.center) to (40.center);
		\draw [thick, in=0, out=-180, looseness=0.75] (43.center) to (26.center);
		\draw [thick] (44.center) to (43.center);
		\draw [thick, <-, in=0, out=180] (39.center) to (44.center);
		\draw [thick] (2.center) to (46.center);
		\draw [thick] (46.center) to (45.center);
		\draw [thick] (45.center) to (3.center);
	\end{pgfonlayer}
\end{tikzpicture}
\end{equation}%
The graph above describes a program that applies an $R_Z$ and an $R_X$ rotation
gate to an input qubit, where the rotation angle is dynamically computed as the
sum of two floats that are given as additional inputs.
The edges in HUGR are statically typed and node operations have a static signature 
(for example, the \texttt{Rz} operation above has the signature $\mathtt{qubit}, \mathtt{f64}
\to \mathtt{qubit}$). We ensure that programs are well-typed by only allowing edges 
that match up with the operation signatures. Note that we leave out type annotations 
in the following examples if they can be inferred from context.

The inputs and outputs of HUGR nodes are explicitly ordered, corresponding to numbered 
\emph{ports} on the nodes. For example, in \Cref{eq:hugr:rz} the \texttt{qubit} 
edge is connected to the first input port of the \texttt{Rz} node, whereas the \texttt{f64} 
edge is connected to its second port. The \texttt{Add} node only has a single output port 
that is wired to both the \texttt{Rz} and \texttt{Rx} node. This is valid since classical 
values can be copied and thus used multiple times.
However, the same is not true for quantum values such as qubits:
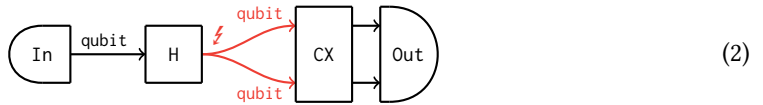
\begin{equation}%
    \label{eq:hugr:nonlinear}
    \begin{tikzpicture}
	\begin{pgfonlayer}{nodelayer}
		\node [style=none] (3) at (1.575, -0.375) {};
		\node [style=none] (4) at (2.325, -0.375) {};
		\node [style=none] (5) at (2.325, 0.375) {};
		\node [style=none] (6) at (1.575, 0.375) {};
		\node [style=none] (7) at (1.95, 0) {\footnotesize \texttt{H}};
		\node [style=none] (15) at (1.025, 0.175) {\scriptsize\texttt{qubit}};
		\node [style=none] (16) at (3.575, -0.625) {};
		\node [style=none] (17) at (4.325, -0.625) {};
		\node [style=none] (18) at (4.325, 0.625) {};
		\node [style=none] (19) at (3.575, 0.625) {};
		\node [style=none] (20) at (3.95, 0) {\footnotesize \texttt{CX}};
		\node [style=none] (21) at (3.575, 0.375) {};
		\node [style=none] (23) at (2.325, 0) {};
		\node [style=none] (39) at (3.575, -0.375) {};
		\node [style=none] (45) at (3.1, 0.525) {\scriptsize\color{ACMRed}{\texttt{qubit}}};
		\node [style=none] (47) at (0.2, 0.375) {};
		\node [style=none] (48) at (0.2, -0.375) {};
		\node [style=none] (49) at (0.2, 0) {\footnotesize \texttt{In}};
		\node [style=none] (50) at (0.575, 0.375) {};
		\node [style=none] (51) at (0.575, -0.375) {};
		\node [style=none] (52) at (0.575, 0) {};
		\node [style=none] (53) at (1.575, 0) {};
		\node [style=none] (54) at (4.325, 0.375) {};
		\node [style=none] (55) at (4.325, -0.375) {};
		\node [style=none] (56) at (4.7, -0.625) {};
		\node [style=none] (57) at (4.825, -0.625) {};
		\node [style=none] (58) at (4.825, 0.625) {};
		\node [style=none] (59) at (4.7, 0.625) {};
		\node [style=none] (60) at (5.075, 0) {\footnotesize \texttt{Out}};
		\node [style=none] (61) at (4.7, 0.375) {};
		\node [style=none] (62) at (4.7, -0.375) {};
		\node [style=none] (63) at (3.1, -0.525) {\scriptsize\color{ACMRed}{\texttt{qubit}}};
		\node [style=none] (64) at (2.55, 0.25) {\color{ACMRed}{\Lightning}};
	\end{pgfonlayer}
	\begin{pgfonlayer}{edgelayer}
		\draw [thick] (6.center) to (3.center);
		\draw [thick] (3.center) to (4.center);
		\draw [thick] (5.center) to (6.center);
		\draw [thick] (16.center) to (17.center);
		\draw [thick] (17.center) to (18.center);
		\draw [thick] (18.center) to (19.center);
		\draw [thick, ->, color=ACMRed, in=180, out=0] (23.center) to (21.center);
		\draw [thick, <-, color=ACMRed, in=0, out=180] (39.center) to (23.center);
		\draw [thick, bend right=90, looseness=2.00] (47.center) to (48.center);
		\draw [thick] (51.center) to (48.center);
		\draw [thick] (51.center) to (50.center);
		\draw [thick] (50.center) to (47.center);
		\draw [thick, ->] (52.center) to (53.center);
		\draw [thick] (59.center) to (56.center);
		\draw [thick] (56.center) to (57.center);
		\draw [thick, bend right=90, looseness=1.75] (57.center) to (58.center);
		\draw [thick] (58.center) to (59.center);
		\draw [thick, ->] (54.center) to (61.center);
		\draw [thick, ->] (55.center) to (62.center);
		\draw [thick] (4.center) to (5.center);
		\draw [thick] (19.center) to (16.center);
	\end{pgfonlayer}
\end{tikzpicture}
\end{equation}
This program would not be physically realisable since the control and target of the
\texttt{CX} gate act on the same qubit. To rule out mistakes like these, HUGR ensures
that ports corresponding to qubits have \emph{exactly one} connected edge, so 
\Cref{eq:hugr:nonlinear} is rejected. This corresponds to treating \texttt{qubit}
as a \emph{linear type}~\cite{Wadler1990LinearTC}.

HUGR continuously enforces these typing and linearity constraints in between optimisation
steps, thus preventing optimisation routines from erroneously invalidating programs.


\subsection{Control Flow \& Hierarchy}%
\label{sec:hugr:hierarchy}

The dataflow representation described above requires additional primitives to
represent control flow in a program. HUGR defines a set of node operations to
express structured control flow.
Given multiple execution graphs, a \texttt{Conditional} operation is able to
branch between them based on a control input:
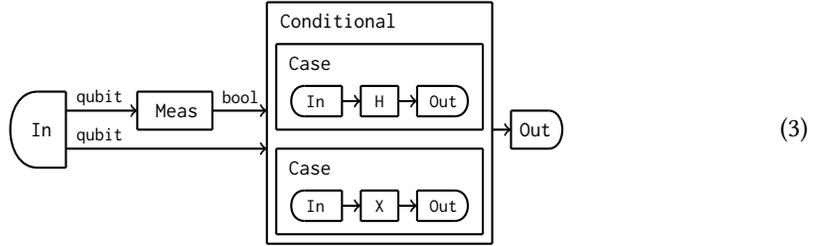
\begin{equation}%
    \label{eq:hugr:conditional}
    \begin{tikzpicture}
	\begin{pgfonlayer}{nodelayer}
		\node [style=none] (0) at (-2.05, 0.5) {};
		\node [style=none] (1) at (-2.05, -0.5) {};
		\node [style=none] (2) at (-2, 0) {\footnotesize\texttt{In}};
		\node [style=none] (10) at (-1.675, -0.25) {};
		\node [style=none] (13) at (-1.225, -0.075) {\scriptsize\texttt{qubit}};
		\node [style=none] (14) at (-1.675, 0.25) {};
		\node [style=none] (15) at (-1.225, 0.425) {\scriptsize\texttt{qubit}};
		\node [style=none] (16) at (-0.725, 0) {};
		\node [style=none] (17) at (0.275, 0) {};
		\node [style=none] (18) at (0.275, 0.5) {};
		\node [style=none] (19) at (-0.725, 0.5) {};
		\node [style=none] (20) at (-0.2, 0.25) {\footnotesize \texttt{Meas}};
		\node [style=none] (22) at (-0.725, 0.25) {};
		\node [style=none] (24) at (0.275, 0.25) {};
		\node [style=none] (41) at (-1.675, -0.5) {};
		\node [style=none] (42) at (-1.675, 0.5) {};
		\node [style=none] (43) at (1, -0.25) {};
		\node [style=none] (44) at (1, -1.5) {};
		\node [style=none] (45) at (4, -1.5) {};
		\node [style=none] (46) at (4, 1.675) {};
		\node [style=none] (47) at (1, 1.675) {};
		\node [style=none] (48) at (1.95, 1.425) {\footnotesize \texttt{Conditional}};
		\node [style=none] (49) at (1, 0.25) {};
		\node [style=none] (50) at (4, 1.125) {};
		\node [style=none] (52) at (1.125, 0) {};
		\node [style=none] (53) at (3.875, 0) {};
		\node [style=none] (54) at (3.875, 1.125) {};
		\node [style=none] (55) at (1.125, 1.125) {};
		\node [style=none] (56) at (1.575, 0.875) {\footnotesize \texttt{Case}};
		\node [style=none] (108) at (0.65, 0.425) {\scriptsize\texttt{bool}};
		\node [style=none] (109) at (1.5, 0.175) {};
		\node [style=none] (110) at (2, 0.175) {};
		\node [style=none] (111) at (2, 0.575) {};
		\node [style=none] (112) at (1.5, 0.575) {};
		\node [style=none] (113) at (1.65, 0.375) {\scriptsize \texttt{In}};
		\node [style=none] (114) at (2, 0.375) {};
		\node [style=none] (115) at (2.25, 0.175) {};
		\node [style=none] (116) at (2.75, 0.175) {};
		\node [style=none] (117) at (2.75, 0.575) {};
		\node [style=none] (118) at (2.25, 0.575) {};
		\node [style=none] (119) at (2.5, 0.375) {\scriptsize \texttt{H}};
		\node [style=none] (120) at (2.25, 0.375) {};
		\node [style=none] (121) at (2.75, 0.375) {};
		\node [style=none] (122) at (3.5, 0.175) {};
		\node [style=none] (123) at (3, 0.175) {};
		\node [style=none] (124) at (3, 0.575) {};
		\node [style=none] (125) at (3.5, 0.575) {};
		\node [style=none] (126) at (3.35, 0.375) {\scriptsize \texttt{Out}};
		\node [style=none] (127) at (3, 0.375) {};
		\node [style=none] (128) at (1.125, -1.375) {};
		\node [style=none] (129) at (3.875, -1.375) {};
		\node [style=none] (130) at (3.875, -0.25) {};
		\node [style=none] (131) at (1.125, -0.25) {};
		\node [style=none] (132) at (1.575, -0.5) {\footnotesize \texttt{Case}};
		\node [style=none] (133) at (1.5, -1.2) {};
		\node [style=none] (134) at (2, -1.2) {};
		\node [style=none] (135) at (2, -0.8) {};
		\node [style=none] (136) at (1.5, -0.8) {};
		\node [style=none] (137) at (1.65, -1) {\scriptsize \texttt{In}};
		\node [style=none] (138) at (2, -1) {};
		\node [style=none] (139) at (2.25, -1.2) {};
		\node [style=none] (140) at (2.75, -1.2) {};
		\node [style=none] (141) at (2.75, -0.8) {};
		\node [style=none] (142) at (2.25, -0.8) {};
		\node [style=none] (143) at (2.5, -1) {\scriptsize \texttt{X}};
		\node [style=none] (144) at (2.25, -1) {};
		\node [style=none] (145) at (2.75, -1) {};
		\node [style=none] (146) at (3.5, -1.2) {};
		\node [style=none] (147) at (3, -1.2) {};
		\node [style=none] (148) at (3, -0.8) {};
		\node [style=none] (149) at (3.5, -0.8) {};
		\node [style=none] (150) at (3.35, -1) {\scriptsize \texttt{Out}};
		\node [style=none] (151) at (3, -1) {};
		\node [style=none] (152) at (4.75, 0.25) {};
		\node [style=none] (153) at (4.75, -0.25) {};
		\node [style=none] (154) at (4.575, 0) {\footnotesize\texttt{Out}};
		\node [style=none] (156) at (4.25, 0) {};
		\node [style=none] (157) at (4.25, -0.25) {};
		\node [style=none] (158) at (4.25, 0.25) {};
		\node [style=none] (159) at (4, 0) {};
	\end{pgfonlayer}
	\begin{pgfonlayer}{edgelayer}
		\draw [thick, bend right=90, looseness=1.25] (0.center) to (1.center);
		\draw [thick] (19.center) to (16.center);
		\draw [thick] (16.center) to (17.center);
		\draw [thick] (17.center) to (18.center);
		\draw [thick] (18.center) to (19.center);
		\draw [thick, ->] (14.center) to (22.center);
		\draw [thick] (0.center) to (42.center);
		\draw [thick] (42.center) to (41.center);
		\draw [thick] (41.center) to (1.center);
		\draw [thick, ->] (10.center) to (43.center);
		\draw [thick] (47.center) to (44.center);
		\draw [thick] (44.center) to (45.center);
		\draw [thick] (45.center) to (46.center);
		\draw [thick] (46.center) to (47.center);
		\draw [thick, ->] (24.center) to (49.center);
		\draw [thick] (55.center) to (52.center);
		\draw [thick] (52.center) to (53.center);
		\draw [thick] (53.center) to (54.center);
		\draw [thick] (54.center) to (55.center);
		\draw [thick, bend right=90, looseness=1.50] (112.center) to (109.center);
		\draw [thick] (109.center) to (110.center);
		\draw [thick] (110.center) to (111.center);
		\draw [thick] (111.center) to (112.center);
		\draw [thick] (118.center) to (115.center);
		\draw [thick] (115.center) to (116.center);
		\draw [thick] (116.center) to (117.center);
		\draw [thick] (117.center) to (118.center);
		\draw [thick, ->] (114.center) to (120.center);
		\draw [thick, bend left=90, looseness=1.50] (125.center) to (122.center);
		\draw [thick] (122.center) to (123.center);
		\draw [thick] (123.center) to (124.center);
		\draw [thick] (124.center) to (125.center);
		\draw [thick, ->] (121.center) to (127.center);
		\draw [thick] (131.center) to (128.center);
		\draw [thick] (128.center) to (129.center);
		\draw [thick] (129.center) to (130.center);
		\draw [thick] (130.center) to (131.center);
		\draw [thick, bend right=90, looseness=1.50] (136.center) to (133.center);
		\draw [thick] (133.center) to (134.center);
		\draw [thick] (134.center) to (135.center);
		\draw [thick] (135.center) to (136.center);
		\draw [thick] (142.center) to (139.center);
		\draw [thick] (139.center) to (140.center);
		\draw [thick] (140.center) to (141.center);
		\draw [thick] (141.center) to (142.center);
		\draw [thick, ->] (138.center) to (144.center);
		\draw [thick, bend left=90, looseness=1.50] (149.center) to (146.center);
		\draw [thick] (146.center) to (147.center);
		\draw [thick] (147.center) to (148.center);
		\draw [thick] (148.center) to (149.center);
		\draw [thick, ->] (145.center) to (151.center);
		\draw [thick, bend left=90, looseness=1.25] (152.center) to (153.center);
		\draw [thick] (152.center) to (158.center);
		\draw [thick] (158.center) to (157.center);
		\draw [thick] (157.center) to (153.center);
		\draw [thick, ->] (159.center) to (156.center);
	\end{pgfonlayer}
\end{tikzpicture}
\end{equation}
Here, the first qubit is measured and depending on the outcome either an
\texttt{H} or an \texttt{X} gate is applied to the second qubit, which is then
outputted. Note that the \texttt{Conditional} node has two \texttt{Case} \emph{child nodes}
that themselves contain children forming a nested dataflow
graph. This highlights another core feature of HUGR: Graphs are
\emph{hierarchical} in the sense that each node can itself contain a nested
child graph. This allows hierarchical nodes like the ones shown in
\Cref{eq:hugr:conditional} to be nested arbitrarily deeply. Besides
\texttt{Conditional}, HUGR also offers a \texttt{TailLoop} primitive to describe
structured looping of a child dataflow graph (see \Cref{sec:examples:tailloop} for an example).

In the spirit of supporting varying levels of abstraction, HUGR also allows
users to specify control flow in non-structured ways via arbitrary control-flow
graphs. Control-flow graphs are expressed via the same
hierarchical structure: \texttt{BasicBlock} nodes contain child graphs
specifying the logic of each block and are then wired together inside a parent
\texttt{CFG} node (see \Cref{sec:examples:cfg} for an example). These graphs can be converted to the aforementioned structured primitives~\cite{Bahmann2015}, or used in lowering stages when targeting CFG based representations like LLVM.


\subsection{Functions and higher order types}%
\label{sec:hugr:higher-order}

Classical programs are often structured as collections of functions in a
namespace, with a defined entry point and internal calls between them.
HUGR provides operations for defining and calling functions, supported by the
hierarchical structure presented in \Cref{sec:hugr:hierarchy}.

Functions can either be defined as a dataflow graph inside a \texttt{FuncDef}
node, or be declared as an external reference with a \texttt{FuncDecl}. In the
latter case, it is assumed that program will be linked with a definition of the
function at a later stage.
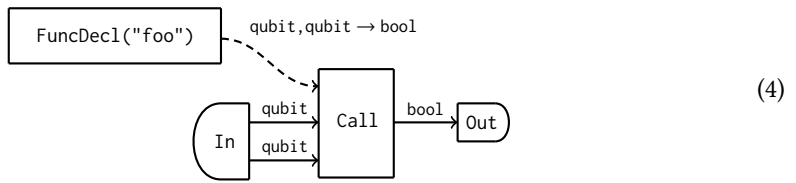
\begin{equation}%
    \label{eq:hugr:func}
    \begin{tikzpicture}
	\begin{pgfonlayer}{nodelayer}
		\node [style=none] (0) at (-2.25, -0.175) {};
		\node [style=none] (1) at (-2.25, -1.175) {};
		\node [style=none] (2) at (-2.2, -0.675) {\footnotesize\texttt{In}};
		\node [style=none] (3) at (-1.875, -0.925) {};
		\node [style=none] (4) at (-1.4, -0.75) {\scriptsize\texttt{qubit}};
		\node [style=none] (5) at (-1.875, -0.425) {};
		\node [style=none] (6) at (-1.4, -0.25) {\scriptsize\texttt{qubit}};
		\node [style=none] (14) at (-1.875, -1.175) {};
		\node [style=none] (15) at (-1.875, -0.175) {};
		\node [style=none] (16) at (-0.95, -0.925) {};
		\node [style=none] (17) at (-0.95, -1.15) {};
		\node [style=none] (18) at (0.05, -1.15) {};
		\node [style=none] (19) at (0.05, 0.275) {};
		\node [style=none] (20) at (-0.95, 0.275) {};
		\node [style=none] (21) at (-0.425, -0.4) {\footnotesize \texttt{Call}};
		\node [style=none] (28) at (-3.65, 0.7) {\footnotesize \texttt{FuncDecl(\textit{"foo"})}};
		\node [style=none] (29) at (0.475, -0.25) {\scriptsize\texttt{bool}};
		\node [style=none] (73) at (1.4, -0.175) {};
		\node [style=none] (74) at (1.4, -0.675) {};
		\node [style=none] (75) at (1.225, -0.425) {\footnotesize\texttt{Out}};
		\node [style=none] (76) at (0.9, -0.425) {};
		\node [style=none] (77) at (0.9, -0.675) {};
		\node [style=none] (78) at (0.9, -0.175) {};
		\node [style=none] (79) at (0.05, -0.425) {};
		\node [style=none] (80) at (-0.75, 0.825) {\scriptsize{\texttt{qubit,qubit} $\to$ \texttt{bool}}};
		\node [style=none] (82) at (-5.075, 1.075) {};
		\node [style=none] (83) at (-2.25, 1.075) {};
		\node [style=none] (84) at (-5.075, 0.35) {};
		\node [style=none] (85) at (-2.25, 0.35) {};
		\node [style=none] (86) at (-2.25, 0.675) {};
		\node [style=none] (87) at (-0.95, -0.425) {};
		\node [style=none] (88) at (-0.95, 0.05) {};
	\end{pgfonlayer}
	\begin{pgfonlayer}{edgelayer}
		\draw [thick, bend right=90, looseness=1.25] (0.center) to (1.center);
		\draw [thick] (0.center) to (15.center);
		\draw [thick] (15.center) to (14.center);
		\draw [thick] (14.center) to (1.center);
		\draw [thick, ->] (3.center) to (16.center);
		\draw [thick] (20.center) to (17.center);
		\draw [thick] (17.center) to (18.center);
		\draw [thick] (18.center) to (19.center);
		\draw [thick] (19.center) to (20.center);
		\draw [thick, bend left=90, looseness=1.25] (73.center) to (74.center);
		\draw [thick] (73.center) to (78.center);
		\draw [thick] (78.center) to (77.center);
		\draw [thick] (77.center) to (74.center);
		\draw [thick, ->] (79.center) to (76.center);
		\draw [thick] (82.center) to (83.center);
		\draw [thick] (82.center) to (84.center);
		\draw [thick] (84.center) to (85.center);
		\draw [thick] (85.center) to (83.center);
		\draw [thick, ->] (5.center) to (87.center);
		\draw [thick, densely dashed, ->, in=180, out=0] (86.center) to (88.center);
	\end{pgfonlayer}
\end{tikzpicture}
\end{equation}
In the example above, we declare an external function \texttt{foo} with
signature $\mathtt{qubit,qubit}\to\mathtt{bool}$ and pass it as an argument to a
\texttt{Call} node, which executes it on some input values.
The wire connecting the declaration to the \texttt{Call} is a special
\emph{constant edge} (represented by a dashed line in \Cref{eq:hugr:func}) that
denotes compile-time static values.
A \texttt{LoadFunction} node may be used to turn such a static function value into a
dynamic runtime value that can be passed around in the dataflow graph.
Combined with the control flow primitives, this allows for the
definition of higher-order functions that take functions as arguments or return
functions as results.



While runtime function values must have a fixed signature, the static function
definitions may have polymorphic signatures. That is, the input and output type
definitions may include type variables that can be instantiated with
user-defined types. The concrete signature for the function is only determined
at the call site.


\subsection{Extensibility}%
\label{sec:hugr:extensibility}

The HUGR representation is designed to be extensible, allowing users to define
new operations and data types specific to their needs.
In the examples presented so far, we have used a set of quantum and classical
operations that are included as part of a standard library for the HUGR
representation. Since the definitions are not hard-coded into the representation,
users are free to mix and replace them with specialised operations relevant to
their domain.
For example, quantum abstractions like quantum control, multiplexed unitaries,
uncomputation, etc. can all be captured inside the extension system and do not
need to be baked into HUGR.

This design choice allows the core HUGR representation to remain agnostic to the
operations being modelled. A program may be defined using the instruction set of
a specific quantum device, and tooling that does not have access to its
definition will still be able to reason about the program structure and perform
optimisations that are agnostic to the operation semantics.
It is the responsibility of the user to implement lowering routines or rewrite
rules that define the behaviour of the new operations, but these are not
required to be shared with the core HUGR implementation.
%


\section{Optimisation}%
\label{sec:optimisation}

The HUGR representation is particularly well suited for pattern-matching based
optimisation. This is a common technique in classical compiler design where
small subgraphs are identified and replaced with more efficient or simpler ones.
In contrast to pass-based optimisation, where the entire program is traversed
and transformed in multiple iterations, pattern matching allows for efficient
composition of rewrite rules and facilitates parallelisation.
%

The port labels on the nodes of a HUGR provide extra structure to the graph
which enables much more efficient matching than generic subgraph isomorphism checks.
Additionally, the incorporation of linear types, which are prevalent in most quantum operations
represented in HUGR, guarantees that the majority of ports have a single
connected edge. Finally, the structured control-flow primitives reduce the
complexity when defining patterns on branching operations. These properties
combined allow us to compile sets of patterns into matching structures which are
able to efficiently search for tens of thousands of patterns
simultaneously~\cite{matching-Mondada2024}.

The operation and type extension framework described in
\Cref{sec:hugr:extensibility} ensures that optimisation routines must always be
aware of potentially unknown operations within the graph. This guarantees that
all rewrite implementations are robust against new operations being added to the
graph and that user-defined routines capable of reasoning about their
domain-specifics can be safely composed with extension-agnostic ones.


\section{Related work}%
\label{sec:discussion}

\label{sec:discussion:related}

\paragraph{Traditional frameworks}
Most traditional quantum compiler frameworks like Cirq~\cite{cirq_developers_2024_11398048}, 
Pennylane~\cite{pennylane}, TKET~\cite{tket}, and Qiskit~\cite{qiskit2024} internally 
represent quantum circuits as lists or graphs of gates and use OpenQASM 2~\cite{qasm2} as a common low-level
assembly format for circuits.
Support for dynamic quantum-classical programs tends to be fairly limited in these frameworks
and usually relies on unrolling of all control-flow.
OpenQASM 3~\cite{qasm3} was introduced to naively handle these classical operations, however it mainly
serves as a high-level programming language rather than an intermediate representation.

\paragraph{QIR}
The Quantum Intermediate Representation (QIR)~\cite{QIRSpec2021} is arguably the most well-know standalone
IR for quantum programs. It is based on the LLVM IR~\cite{llvm} and leverages the existing
mature compiler infrastructure of the LLVM project.
QIR is designed to be hardware-agnostic and as such offers a notion of \emph{profiles} to
specify the capabilities offered by a given device.
In particular, there is ongoing work~\cite{Campora2023} to define a profile for QIR programs that
captures the real-time classical operations and branching demonstrated 
in~\cite{lubinski2022advancing, brown2023advancescompilationquantumhardware}.

Compared to HUGR, QIR is a more low-level representation where qubits are treated like opaque
pointers and quantum operations are side-effectful opaque functions.
This means that optimisers like that in~\cite{luo2024dataflowbasedoptimizationquantumintermediate}
need to rely on global dataflow analyses to track
qubits as opposed to the simpler graph-based matching available in HUGR (see \Cref{sec:optimisation}).
Furthermore, QIR is not easily extensible:
while it provides some built-in quantum features like controlled and
adjoint operations, there is currently no way to add custom higher-level
abstractions.

\paragraph{MLIR}
QIR's lack of customisability is at least in part due to the rigidness of LLVM's IR which was mainly designed for C-like languages.
The MLIR project~\cite{mlir} aims to address LLVM's drawbacks by introducing an IR with a dialect system that allows users to define their own domain-specific abstractions.
This design served as a major inspiration for the extension system in HUGR.

While MLIR was initially used in the context of heterogeneous computing and machine learning,
it has since also been applied to the quantum domain.
MLIR based quantum dialects are used in QIRO~\cite{qiro}, QCor~\cite{McCaskey2021MlirDialect}, and the Catalyst compiler~\cite{catalyst}.
In fact, MLIR is expressive enough such that HUGR itself can be implemented as one of its dialects, which would make it compatible with the broader MLIR ecosystem.
A prototype of such a dialect with conversions to and from the reference implementation
has already been developed.\footnote{Available at \href{http://www.github.com/CQCL/hugr-mlir}{\texttt{github.com/CQCL/hugr-mlir}}.}

However, we decided to keep the reference implementation of HUGR independent of MLIR for a few key reasons.
First, MLIR is still under rapid development, and as such not fully stable and mature.
Furthermore, as part of our mission-critical stack, we want to leverage the memory safety provided by Rust and avoid being tied to MLIR's C++ implementation.
Finally, maintaining our own implementation allows us to focus on features that are particularly relevant to the quantum domain.
For example, linear types are a first-class concept in HUGR whereas representing them in MLIR would require a more complicated setup and checks that would feel less natural.

\bibliographystyle{plainurl}
\bibliography{bibliography}


\newpage
\onecolumn
\begin{appendix}

\section{Additional Examples}%
\label{sec:examples}

\subsection{Tail Loops}%
\label{sec:examples:tailloop}

Below is an example HUGR program that uses a \texttt{TailLoop} node to implement a $(I + i\sqrt{2}X)/\sqrt{3}$
operation on its input qubit using the repeat-until-success scheme from \cite[Fig. 8]{rus-paetznick2013}.
The first \texttt{bool} output of the loop body controls whether another loop iteration should be performed,
which is the case if the measurement returned \texttt{false}.
The \texttt{Conditional} node is used to apply an additional \texttt{Z} correction in that case.
Note that the \texttt{T}$_{\text{\ttfamily X}}^\dagger$ nodes stand for the $HT^\dagger H \approx R_X(-\pi/4)$ gate.
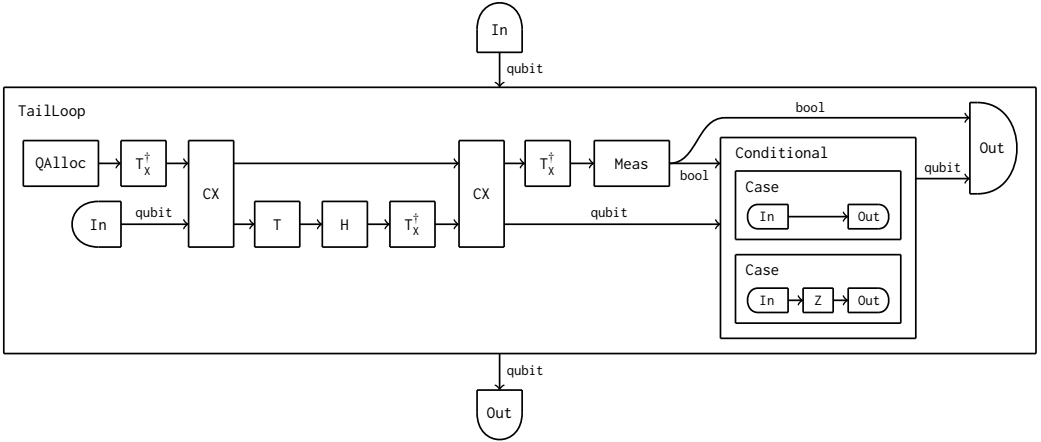
\[ \scalebox{0.8}{\input{\tikzitpathfigures/loop.tikz}} \]

\newpage

\subsection{Control Flow Graphs}%
\label{sec:examples:cfg}

The graph below implements the same program as in \Cref{sec:examples:tailloop} using
a control-flow graph instead of a \texttt{TailLoop} node.
The dotted edges between the basic blocks describe control flow instead of dataflow and thus
are not required to be acyclic.
\[ \input{\tikzitpathfigures/cfg.tikz} \]

\end{appendix}

\end{document}